\documentclass{article}

\usepackage{arxiv}

\usepackage[utf8]{inputenc} % allow utf-8 input
\usepackage[T1]{fontenc}    % use 8-bit T1 fonts
\usepackage{hyperref}       % hyperlinks
\usepackage{url}            % simple URL typesetting
\usepackage{booktabs}       % professional-quality tables
\usepackage{amsfonts}       % blackboard math symbols
\usepackage{nicefrac}       % compact symbols for 1/2, etc.
\usepackage{microtype}      % microtypography
\usepackage{lipsum}
\usepackage{graphicx}
\usepackage{multirow}

\usepackage{listings}
\usepackage{xcolor}
\usepackage{caption}
\usepackage{subcaption}
\usepackage{booktabs}
\usepackage{varwidth}
\usepackage{array}  % for "\extrarowheight" macro

\graphicspath{ {./figures/} }

\title{PhyAAt: \textbf{Phy}siology of \textbf{A}uditory \textbf{At}tention to Speech Dataset}

\date{}
\author{
  Nikesh~Bajaj$^{a,b}$, Jes\'us~Requena~Carri\'on$^{a}$, Francesco Bellotti$^{b}$\\
  $^{a}$Centre for Intelligent Sensing, School of Electronics Engineering and Computer Science (EECS),\\
  Queen Mary University of London, London, UK\\
  $^{b}$Elios Lab, Dipartimento di Ingegneria Navale, Elettrica, Elettronica e delle\\ Telecomunicazioni (DITEN), University of Genoa, Italy\\
  \texttt{\{n.bajaj, j.requena\}@qmul.ac.uk, franz@elios.unige.it}\\
}

\begin{document}

\maketitle

\begin{abstract}
Auditory attention to natural speech is a complex brain process. Its quantification from physiological signals can be valuable to improving and widening the range of applications of current brain-computer-interface systems, however it remains a challenging task. In this article, we present a dataset of physiological signals collected from an experiment on auditory attention to natural speech. In this experiment, auditory stimuli consisting of reproductions of English sentences in different auditory conditions were presented to 25 non-native participants, who were asked to transcribe the sentences. During the experiment, 14 channel electroencephalogram, galvanic skin response, and photoplethysmogram signals were collected from each participant. Based on the number of correctly transcribed words, an attention score was obtained for each auditory stimulus presented to subjects. A strong correlation ($p<<0.0001$) between the attention score and the auditory conditions was found. We also formulate four different predictive tasks involving the collected dataset and develop a feature extraction framework. The results for each predictive task are obtained using a Support Vector Machine with spectral features, and are better than chance level. The dataset has been made publicly available for further research, along with the python library \textit{phyaat} to facilitate the preprocessing, modeling, and reproduction of the results presented in this paper. The dataset and other resources are shared on webpage - \href{https://phyaat.github.io}{{\it https://phyaat.github.io}}.

%The dataset and 

%The correlation analysis presented shows that auditory attention is very much depended on the auditory conditions. 
%We formulated four different predictive tasks for collected dataset and explain the feature extraction approaches, keeping the basic mechanism of auditory attention in mind. 

%All the results, codes and tutorial are updated regularly on - \href{https://PhyAAt.github.io}{https://PhyAAt.github.io}.

%By using correlation analysis, the effects of the auditory condition, the type of task and the attention score on different spectral EEG bands was assessed. 

%We formulated four predictive problems using the spectrum of recorded EEG signals as predictors and approached them with the state-of-the-art machine learning algorithms, including support vector machines and decision trees. Our results indicate that the analysis of EEG signals can reveal details about the state of the auditory attention system. Our dataset offers the unique opportunity to explore how auditory attention manifests in physiological signals and it has been made publicly available to support research on human natural speech processing and auditory attention. 
\end{abstract}

% keywords can be removed

\section{Introduction}
\label{S:Intro}
%\vspace{3.5cm}
Auditory attention is a cognitive process of great importance that still remains poorly understood \cite{AudAtten2000}. It plays a vital role in many human activities, from aircraft flying to learning in a classroom environment. Thus, understanding attention can be important to improve the effectiveness, experience, and performance of many tasks \cite{SelAtte2001}. 
A better understanding of auditory attention could also lead to improved brain-computer-interface (BCI) systems that use emotional and mental states to execute actions.

Auditory attention and auditory perception are closely related processes; while perception is the process of interpreting information, attention is the process by which discrete pieces of information are selected or discarded \cite{styles2004attention}. Auditory perception is a complex brain process that involves encoding sounds into patterns of neural activity \cite{theunissen2004methods}. This process can also modulate other physiological responses, such as the heart rate and the sympathetic tone. The brain activity can be investigated in the Electroencephalogram (EEG), and the heart rate and sympathetic response can be extracted from Photoplethysmogram (PPG) and Galvanic Skin Response (GSR) signals, respectively. Other physiological modalities such as the Electromyogram, the Electrooculogram, and the Electrocardiogram are also available. 

Physiological signals are used in many applications from healthcare to BCI system. Datasets of physiological signals are available for a wide variety of applications, from paralysis \cite{BCIPararylysis2008} and epilepsy \cite{andrzejak2012nonrandomness} studies, to applications for patients in a complete locked-in state \cite{BCISpeller2011, BCIAud2012, BCILockedin2011} to emotion recognition \cite{EmoReco2004, EmoPhys2005, EmoEEG2006, koelstra2012deap, EEGMusic2010} and % listening to music \cite{EEGMusic2010}, 
serious games \cite{EmoGame2011, EEGgame2010, EEGgame2011}. Widely used EEG datasets for BCI applications include motor imagery datasets \cite{brunner2008bci, leeb2007brain, shin2017open}. There are several attention related datasets, including datasets for covert and overt visual attention \cite{aloise2012covert, arico2014influence, treder2011brain} and for auditory-visual attention shift \cite{vceponiene2008modality}. Specifically for auditory attention based on EEG signals, an auditory oddball paradigm has been presented, where the response of participants to oddball sounds inserted in streams of sinusoidal tones was analysed \cite{halder2010auditory, schreuder2010new}.

Among the wide variety of physiological datasets available, to the best of our knowledge there is a lack of datasets designed to study auditory attention to natural speech. With the goal of supporting future research in the field of auditory attention, we have designed an experiment based on the dichotic listening task \cite{dichoticTask2008}, and collected a dataset of physiological signals that include 14 channel EEG, PPG and GSR. % with 25 subjects. 
The dataset is available to the wider community for it to be used freely\footnote{https://phyaat.github.io}. Based on this dataset, a model could be trained to predict from physiological signals the attention level of individuals in different scenarios, for instance during the delivery of a lecture. %physiological responses can be used to analyse the content of speech and delivery of lectures to evaluate the quality. 
This dataset does not restrict itself to auditory attention studies and can also be used to investigate other mechanisms involved in brain auditory processing of natural speech. One example of such applications %with analysis of attention score from collected dataset, was 
is the design of the difficulty levels of a game based on the level of auditory attention \cite{bajaj2018auditory}. 

The rest of the article is organised as follows. Section \ref{S:Exp} explains the experimental design, materials and procedures. Section \ref{S:Dataset} describes the collected dataset, labels and file structure. In Section \ref{S:AAtScore}, we analyse the correlation between attention score and auditory conditions. In Section \ref{S:pTasks}, we formulate the four predictive tasks and Section \ref{S:fExtraction} describes the framework to extract the features from physiological responses. Section \ref{S:Results} demonstrates the results of the predictive tasks using the data from one subject. Section \ref{S:Conclusion} concludes the presented work and discusses several directions for future research. The details of resources made available are explained in the Resources section at the end of the paper. 

%need and applications

\section{Experiment}%paradigm
\label{S:Exp}
\begin{figure}[t]
    \centering
    \includegraphics[width=0.99\textwidth]{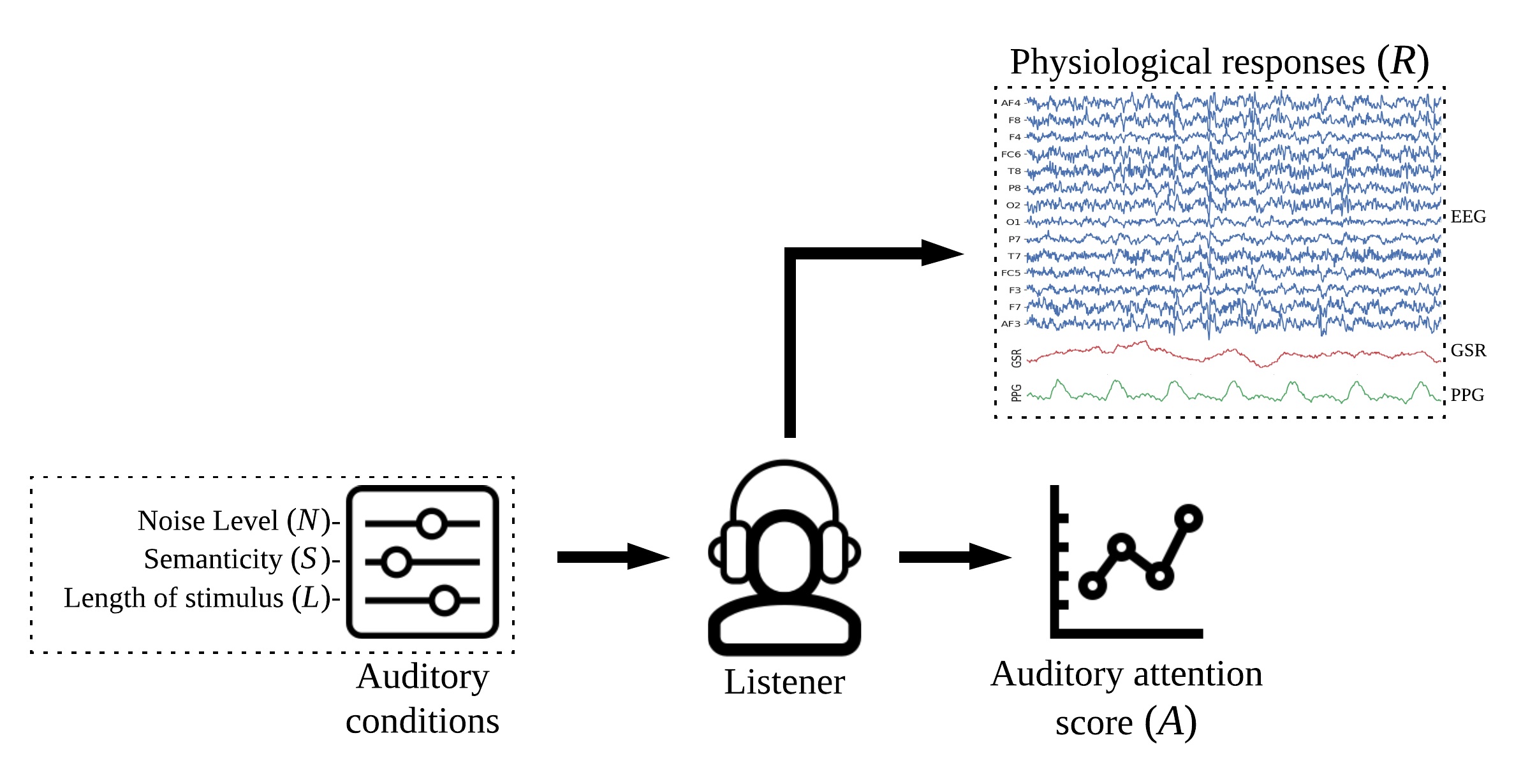}
    \caption{Experiment Design}
    \label{fig:ED}
\end{figure}

\subsection{Overview of the experiment design}
\label{ss:ExDesign}
%%% - Overall experiment design setup at high level {given the definition of diochotic listening task}
%To collect the dataset for auditory attention, 

An experiment based on diochotic listening task was designed to create the dataset for auditory attention analysis. The main components of this experiment are shown in Figure \ref{fig:ED}. Unlike the diochotic listening task, only one audio stimuli per trial was presented to a participant. Audio stimuli were reproduced under three different auditory conditions ($N$,$S$, and $L$) and following previous studies, the attention score $A$ for each trial was computed from a transcribed audio message, by counting the number of correctly identified words \cite{auralPerception1967, dichoticTask1987, dichoticTask2007, dichoticTask2003}. While the establishing attention score, minor spelling errors and typos were ignored such as {\it did/dids}, {\it beautiful/beutiful}, {\it dog/dogs}. For the entire duration of the experiment, three different modalities of physiological responses ($R$) were recorded at the sampling rate of 128Hz. The details of the methods and materials are explained in subsequent subsections.
%%% - Specific details about - stimuli, participants, signals, procedure, trials

%%% -  Stimuli, (S) semantic - non-semantic, (N) noise level, (L) length 
\subsection{Stimuli}
\label{ss:stumli}
A total collection of 5000 audio files, each containing semantically correct English language sentence was obtained from the Tatoeba Project \cite{tatoeba}. The length of stimuli ranges from 3 to 13 words per sentence. All the stimuli were reproduced by the same speaker to avoid variations in physical properties such as pitch, rate of speech, etc. From the collection of semantic stimuli, a total 1700 non-semantic stimuli were created by suitably inserting random isolated words in between. The following are examples of semantic and non-semantic stimuli:
\begin{eqnarray*}
&\textbf{Tx1: } \textit{I am going to study.}\\
&\textbf{Tx2: } \textit{I would like to read some books about the Beatles.}\\
&\textbf{Tx3: } \textit{Let's \textbf{touch enjoyable} go.}\\
&\textbf{Tx4: } \textit{I have  a \textbf{hey} big \textbf{are we} dog.}
\end{eqnarray*}
In the above examples, stimuli \textbf{Tx1} and \textbf{Tx2} are semantic and stimuli \textbf{Tx3} and \textbf{Tx4} are non-semantic stimuli generated by inserting the words highlighted in bold font. From each group of semantic and non-semantic stimuli, six additional groups of stimuli were created by adding different levels of background noise. The signal to noise ratios (SNR) of each background noise level were -6 dB, -3 dB, 0 dB, 3 dB, 6 dB and noise-free ($\infty$ dB). In summary, we created two groups of audio stimuli, semantic and non-semantic, of lengths ranging from 3 to 13 words and six different levels of background noise was added to each audio stimuli during reproduction.

%% - should this be included??????
%Stimuli (semantic and non-semantic) were also grouped into three categories, namely small (L1), medium (L2) and long (L3), depending on the number of words contained. The average number of words in L1, L2 and L3 are 4, 8, and 12, with a variation of one word, respectively. This categorization is close to the one presented in \cite{auralPerception1967}. The variation of one word was allowed, assuming, it has no effect on perception, as studies \cite{primacyModel1998, WordLengthEffects1995} have shown.
%---------------------

%\textcolor{red}{Length L1, L2, L3?}
%According to length, each stimuli from above groups, is categorised into three different sub-groups, namely; small (L1), medium (L2), and large (L3), with length of 3 to 5 words, 7 to 9 words, and 11 to 13 words, respectively.   

%%% - $R$ - given the acronym (EEG, GSR, PPG), recording HW, SW, processing, sampling rate
\subsection{Physiological signals}
We recorded three types of physiological responses from each participant, namely EEG, GSR, and PPG. A 14 channel Emotiv Epoc wireless device was used for recording EEG signals \cite{EmotivEpoc}. The EEG signals were collected using the API provided by the device manufacturer with the highest available sampling rate of 128Hz. The DC component was removed from EEG signals while recording. Two small copper plates placed on the index finger and middle finger were used to measure the GSR. The resulting GSR signal was low-pass filtered and recorded at a rate of 128 Hz. Both the raw and low-pass filtered signals were recorded. 

A pulse sensor \cite{PPG} positioned on the ring finger was used to record the PPG response, which measures microvascular blood volume changes. Pulse rate and the inter-beat interval (IBI) were obtained from the PPG response using software code provided by the sensor manufacturer \cite{PPGCode}. 

In summary, 19 signals streams were recorded from each participant, i.e. 14 EEG channels, 2 GSR streams (raw and low-pass filtered) and 3 PPG streams (raw signal, pulse rate, and IBI). All the physiological responses were labeled accordingly.  

%%% - Participants, First Language, age-group, right-left handed, demographic
\subsection{Participants}
\label{ss:participants}
A total of 25 university students, 21 male, 4 female, 1 left handed, 24 right handed,  participated in our study. All the participants were non-native English speakers (i.e. their first language was other than English) and had no known auditory processing disorder. The age of the participants ranged from 16 to 34 years. The majority of nationalities among the participants were Indian and Italian and the majority of the first languages were Arabic, Italian and Malayalam. 

%%% - time line, trial details, stimuli selection - case generation, 
\subsection{Experiment procedure}
\label{ss:Procedure}
After mounting all the sensors e.g. EEG headset, GSR plates, and PPG pulse sensor, participant were provided with a passive ear-phone and presented with a computer interface. Participants were then asked to enter demographic information that included nationality, gender, age-group, and first language. Participants were also asked to rate their English language skills in terms of listening, writing, reading, and speaking. Once all the information was entered, participants could initiate the experiment. 

A total of 144 stimuli were randomly selected without replacement, for each participant. Out of 144, 72 were selected from the semantic group and 72 from the non-semantic group. In each group, the 72 stimuli were equally split into six background noise levels, i.e. 12 stimuli were assigned to each noise level. Furthermore, the 12 stimuli within each noise level had different lengths. Specifically, 5 stimuli were short with an average of 4 words (L1), 4 stimuli had an average of 8 words (L2), and 3 stimuli were long, with an average of 12 words (L3). The chosen lengths followed closely previous studies \cite{auralPerception1967}. 
%stimuli presented different lengths for a noise level are chosen with three different length group, namely; small (L1), medium (L2), and long (L3). The distribution of 12 stimuli for each length is not equal, i.e. 5 stimuli from L1, 4 from L2, and 3 from L3. The average length of stimuli for L1 is 4 words, for L2 is 8, and for L3 is 12, 
The variation in length within each category was $\pm1$ word and we assumed that this variation did not have a significant effect on perception, as suggested by previous studies \cite{primacyModel1998, WordLengthEffects1995}.

\begin{figure}[t]
     \centering
     \begin{subfigure}[b]{0.40\textwidth}
         \centering
         \includegraphics[width=\textwidth]{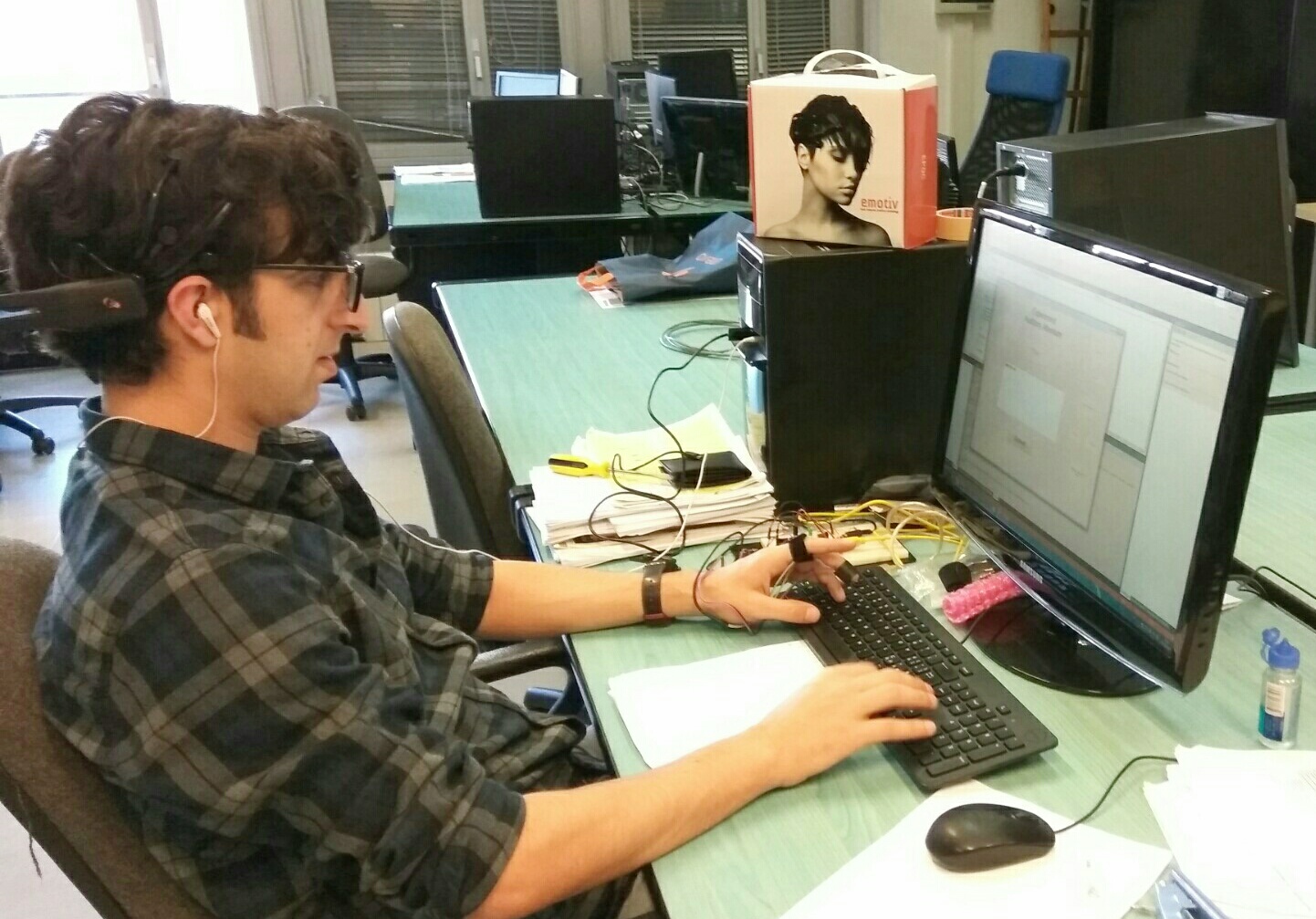}
         \caption{A participant during the experiment}
         \label{fig:ex_participant}
     \end{subfigure}
     \hfill
     \begin{subfigure}[b]{0.53\textwidth}
         \centering
         \includegraphics[width=\textwidth]{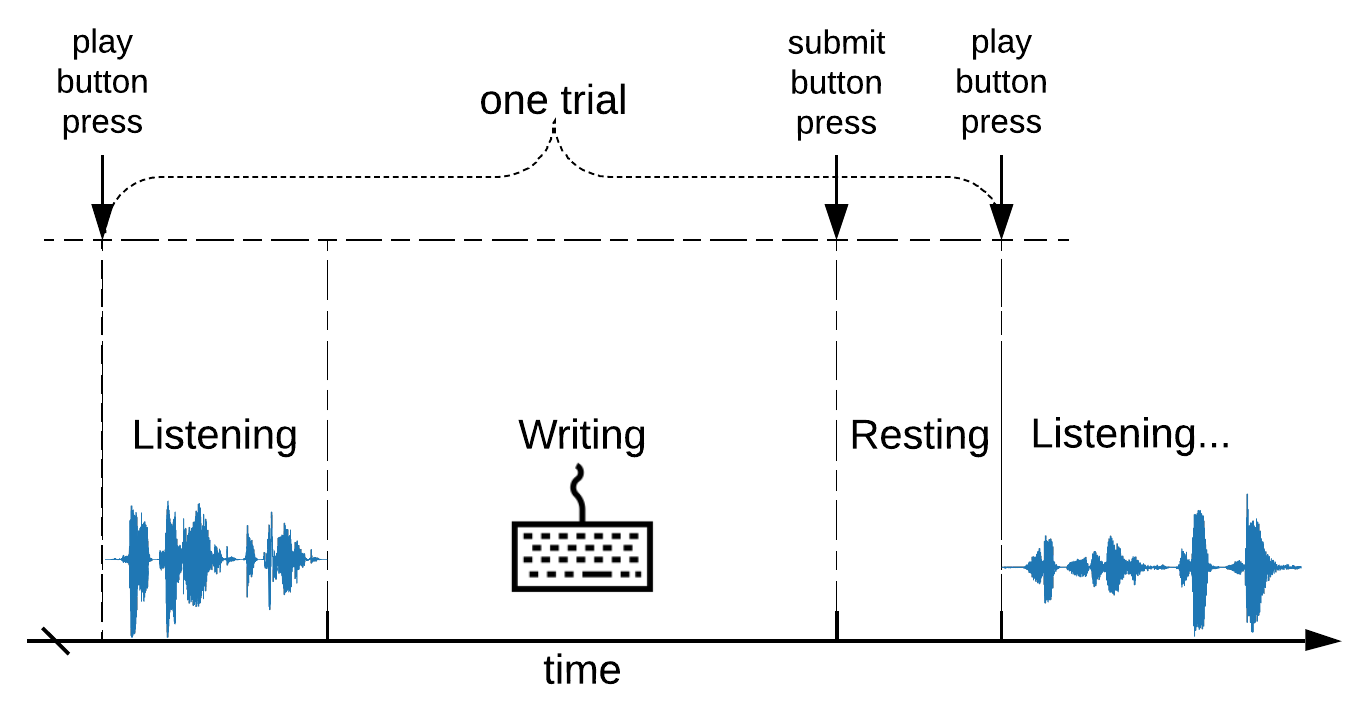}
         \caption{Timeline of a trial}
         \label{fig:ex_timeline}
     \end{subfigure}
        \caption{Experiment procedure}
        \label{fig:ex_procedure}
\end{figure}

The experiment thus consisted of 144 trials per participant, during which they had to transcribe one auditory stimulus. Each trial was divided into three tasks, namely \textit{Listening, Writing,} and \textit{Resting}. The timeline of a trial is shown in Figure \ref{fig:ex_timeline}. A participant could actively start a trial by pressing the play button on the computer interface, after which a listening task would start by reproducing one of the 144 audio stimuli. Once the audio was finished, participants were allowed to write the response in a text-box. Participants were not allowed to replay any stimuli and the interface remained disabled while stimuli were being played. After the transcription was written, participants could press the submit button to save the response and end the writing task. To start new trial, participants had to press the play button again. The time period between writing and the next listening task was labeled as resting. 

Participant were explained the entire procedure beforehand. The average time taken for a participant to complete the experiment was $40\pm10$ minutes. Figure \ref{fig:ex_participant} shows one of the participants conducting the experiment.

\section{Collected dataset}

\label{S:Dataset}
\begin{figure}[ht]
 \centering
    \includegraphics[width=0.88\textwidth]{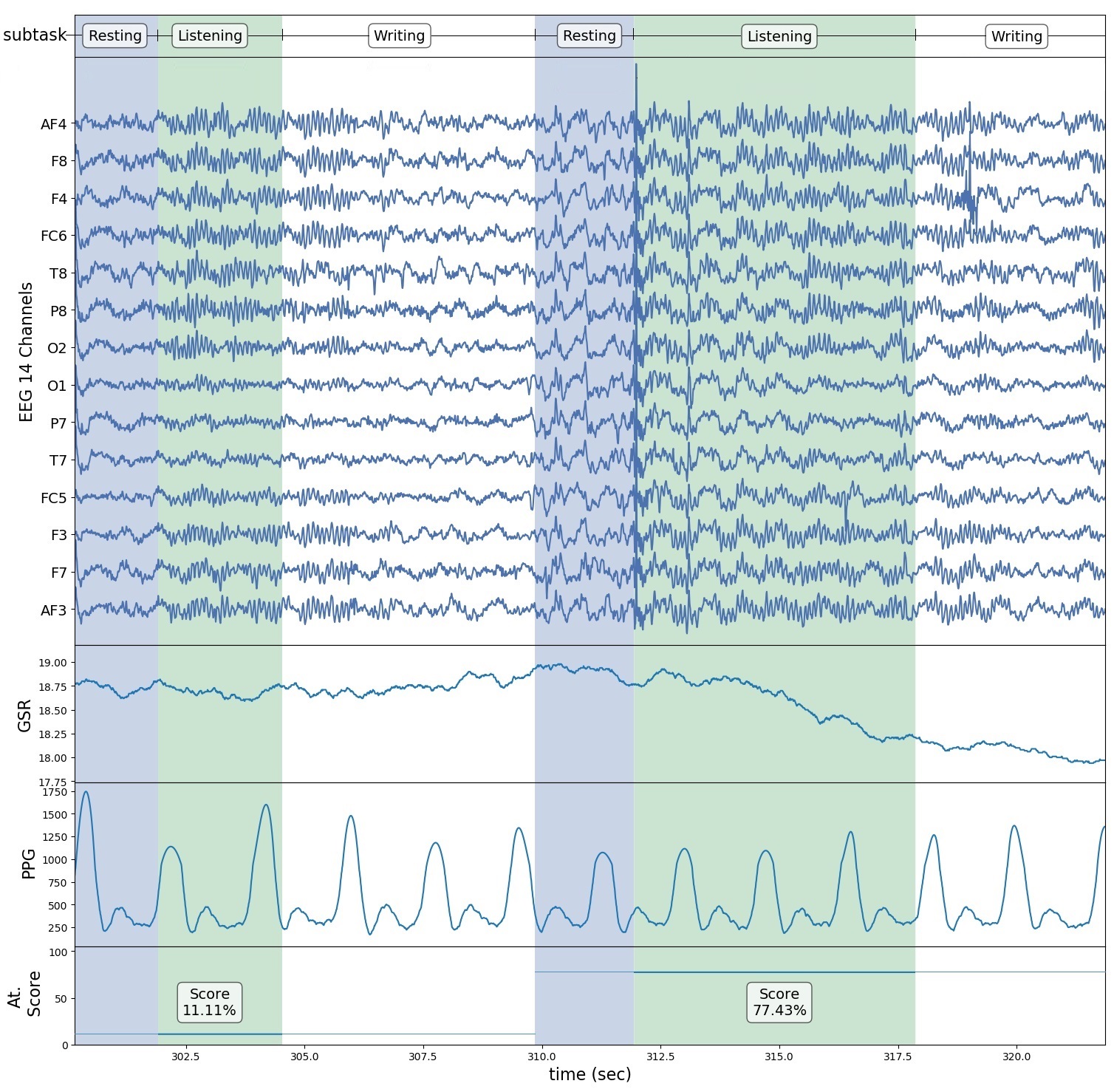}
    \caption{A segment of collected physiological responses with corresponding labels. Green color background is corresponds to listening task, white for writing, and blue for resting.}
    \label{fig:signals}
\end{figure}

The collected dataset includes the physiological responses of the 25 participants. Signals and time segments are conveniently labeled so as to identify auditory conditions and tasks, and the computed attention scores for each stimuli is also included. A segment of physiological responses and corresponding labels is shown in Figure \ref{fig:signals} and includes the 14 EEG channels, the low-pass GSR signal and the raw PPG response. Listening, writing, and resting segments are identified by different background colors (green, white, and blue, respectively) and the attention score for listening segments is also shown. A summary of collected dataset is described in Table \ref{table:Summary}.

As shown in Figure \ref{fig:file_stcr}, the collected dataset is structured in a directory containing 25 sub-directories, one for each participant, named as \textit{S1, S2, ... S25}. Each sub-directory contains two files namely \textit{S[x]\_Signals.csv} and \textit{S[x]\_Textscores.csv}, where {\it [x]} is the participant ID, e.g. {\it S1} corresponds to the participant 1. The signal file includes 19 streams of signals, time-stamped, and three labels, namely \textit{Label\_N, Label\_S}, and \textit{Label\_T}, corresponding to noise level, semanticity, and task respectively. For the noise level $\infty$ dB, the numerical value 1000 is used. Semanticity is encoded as 0 and 1, for semantic and non-semantic respectively. Tasks are encoded as 0, 1, and 2 for listening, writing, and resting respectively. The labels before first trial are set to -1. The text score file includes the attention score (\textit{Label\_A}) for each trial, along with noise level, semanticity, length of stimulus, and timestamp. The dataset has been anonymised and the time-stamp in each files are set to the same time, i.e. 00 hours. The dataset also contains the demographic information and English level self-rating score of each participant. Finally, a {\it ReadMe} file is included. 

A python library \textit{phyaat} has been created to allow users to download the dataset and extract the segments with the corresponding labels to use for predictive modeling. User guide and instructions are also provided \footnote{\url{https://phyaat.github.io/introduction}}.

\begin{table}[ht]
\begin{minipage}[b]{0.55\linewidth}
\centering
    {\renewcommand{\arraystretch}{1.3}
    \centering
    \caption{Dataset summary}
    %\begin{tabular}{l | l}
    \begin{tabular}{p{3cm}|p{6cm}}
    \hline
    %\textbf{Signal} & \textbf{Specifications}\\
    \hline
    Types of signal &  EEG, GSR, PPG\\
    EEG channels & 14 channels;  AF3, AF4, F3, F4, FC5, FC6, F7, F8, T7, T8, P7, P8, O1, O2\\
    Total signals streams & 19 streams;  14 from EEG, 2 from GSR- instant sample and moving averaged, 3 from PPG - raw signal, beats per minutes (BPM), interval between beats (IBI)\\
    Sampling rate & 128Hz \\
    Participants & 25 participants; 21 male, 4 female \\
    Number of stimuli per participants & 144 randomly selected stimuli, 72 semantic and 72 non-semantic\\
    Independent variables & Noise level, Semanticity, Length of stimulus\\
    Noise levels & 6 levels; -6, -3, 0, 3, 6 and $\infty$ dB SNR\\
    Semanticity levels & 2 levels; 0-semantic, 1-non-semantic\\
    %Length of stimulus & L1(3-5), L2(7-9) and L3(11-13) words\\
    Length of stimulus & 3 to 13 words per stimulus, grouped into three categories; L1 (small), L2 (medium) and L3 (long) \\
    Average duration of a stimuli & $3 (\pm1.2)$ sec \\
    Average duration of entire recording of a participant & $40 (\pm10)$ mins\\
    Self-rating & Writing, listening, reading and speaking skill of English language, rating scale 1-5\\
    %Total number of responses & $144\times25 = 3600$\\
    \hline
    \hline
    \end{tabular}
    \label{table:Summary}
    }
\end{minipage}\hfill
\begin{minipage}[b]{0.4\linewidth}
    \centering
    \includegraphics[height=1.2\textwidth]{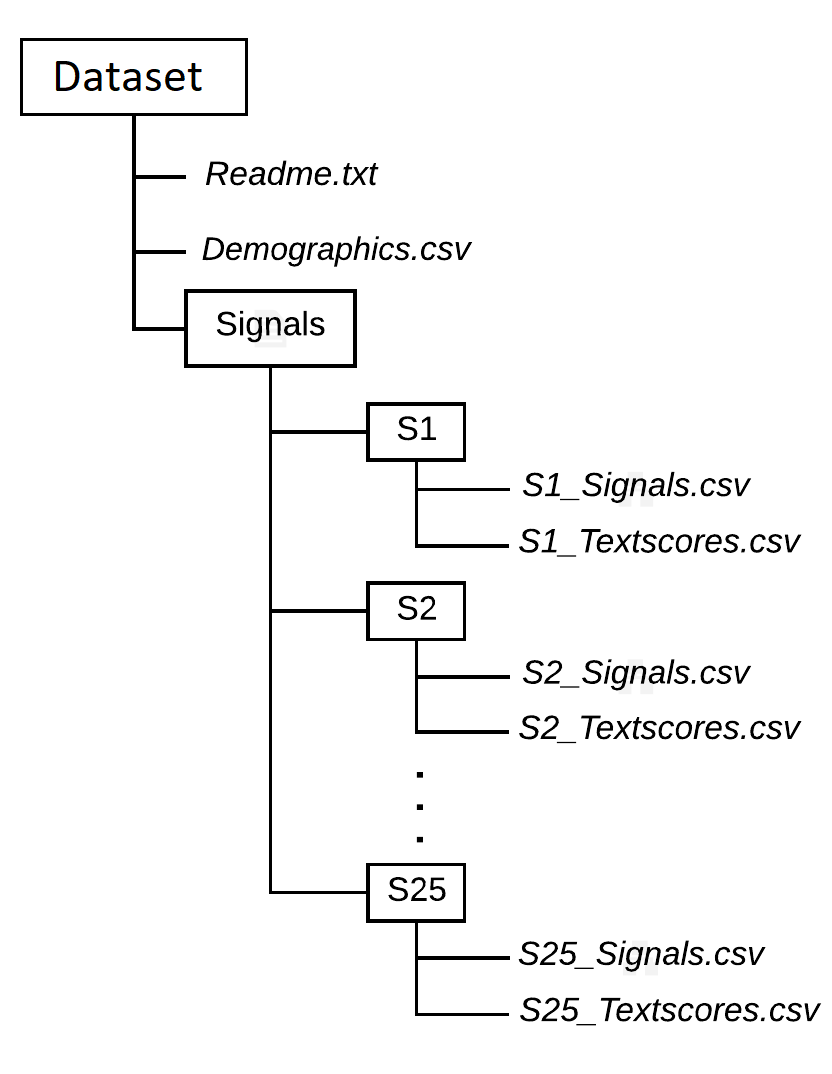}
    \vspace*{.6in}
    \captionof{figure}{File structure of dataset directory}
    \label{fig:file_stcr}
\end{minipage}
\end{table}

\section{Analysis of Attention Score}
\label{S:AAtScore}
\begin{figure}[ht]
     \centering
     \begin{subfigure}[b]{0.39\textwidth}
         \centering
         \includegraphics[width=\textwidth]{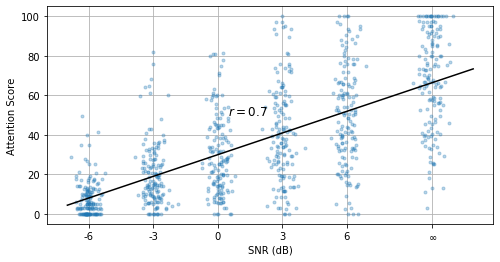}
         \caption{Noise level}
         \label{fig:att_noise}
     \end{subfigure}
     \hfill
     \begin{subfigure}[b]{0.3\textwidth}
         \centering
         \includegraphics[width=\textwidth]{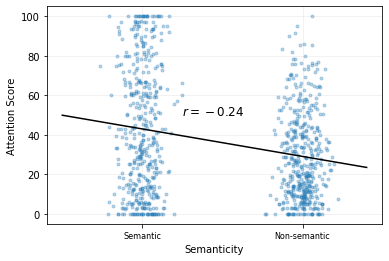}
         \caption{Semanticity}
         \label{fig:att_sem}
     \end{subfigure}
     \hfill
     \begin{subfigure}[b]{0.3\textwidth}
         \centering
         \includegraphics[width=\textwidth]{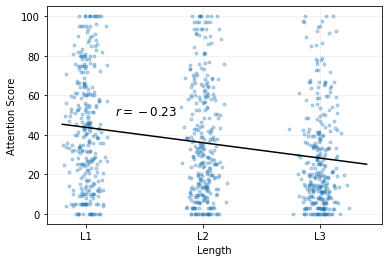}
         \caption{Length of stimulus}
         \label{fig:att_len}
     \end{subfigure}
    \caption{Correlation of attention score and auditory conditions, $p<0.0001$}
    \label{fig:att_score}
\end{figure}

The Spearman rank correlation between the attention score and the auditory condition (i.e. noise level, semanticity, and length of stimulus) was carried out and its results are shown in Figure \ref{fig:att_score}. It can be observed that attention score has a strong positive correlation ($r=0.7$) with the background noise level and a negative correlation with the semancity ($r=-0.24$) and length of the stimulus ($r=-0.23$), all with a p-value $p<0.0001$. Furthermore, a repeated-measure ANOVA test was performed on the attention score considering each auditory conditions individually and jointly. The p-values were found to be low ($p<<0.001$), suggesting a significant impact of the auditory conditions on the attention scores.

\section{Formulation of predictive tasks}
\label{S:pTasks}
Based on the analysis of the attention score and the experiment design (Figure \ref{fig:ED}), four different predictive tasks have been formulated. %In this section we use the notation $Sg_L$, $Sg_W$ and $Sg_R$ for the listening, writing and resting segments respectively, extracted from the physiological responses, denoted collectively by $R$. Each segment contains all the physiological signals (EEG, GSR, PPG) that were recorded.

\subsection{\textbf{T1}: Attention score prediction}
\label{ss:T1}

In this task, the objective is to estimate the auditory attention level $A$ of a participant based on the physiological signals recorded during the listening segment. %$Sg_L$. %, i.e. $A \approx f(Sg_L)$. 
%from physiological signals $R$, assuming $A \approx f(R)$. Since the attention score is respect to listening task only, the attention score can be considered to be a function of physiological responses from listening segment, $A \approx f(Sg_L)$ for $R \rightarrow Sg_L$. 
Given a set of features $F_r$ extracted from the listening segment %$Sg_L$, 
this objective can be formulated as finding a model $f_A(\cdot)$ such that the attention score can be approximated by the quantity
\begin{equation}
A^{\prime} = f_A(F_r)
\end{equation}
%For a predictive modeling of attention score $A$, a set of features can be extracted from listening segment i.e. $Sg_L \rightarrow F_r$ and a function $f(\cdot)$ can be modeled with corresponding attention score $A$, such that the attention score for test data (unseen listening segments), $A^{\prime}$, can be estimated as;
%$$A^{\prime} = f(F_r)$$
The model $f_A(\cdot)$ is defined as the one minimising the expected risk 
\begin{equation}
\mathcal{E}(f_A) = \mathbb{E}[\mathcal{L}(A,f_A(F_r))],
\end{equation}
where $\mathcal{L}(\cdot,\cdot)$ is the chosen loss function and $\mathbb{E}[\cdot]$ is the expectation operator. %As the attention score ranges from 0 to 100, 
The choice of the loss function can be defined as the mean square error, the mean absolute error or a combination of the two, such as the Huber loss.

%$$A \approx f(R).$$
%$$\mathcal{A} \approx f(\mathcal{R}).$$
%$$\mathpzc{A} \approx f(\mathpzc{R}).$$
%$$\varmathbb{A} \approx f(\varmathbbc{R}).$$
%$$\symA \approx f(\symC).$$

The analysis of the attention score (c.f. Section \ref{S:AAtScore}) shows that there is a strong correlation between attention score and auditory conditions, such as noise level and semanticity. %, i.e. $A = f(N, S, L)$. 
Therefore our knowledge about the auditory conditions can be informative for predicting the auditory attention. Although in general auditory conditions are not known, in our controlled experiment we do have access to their exact values. Consequently, we can formulate a predictive task for the auditory attention, that use the value of the auditory conditions as input.

%Thus, if we assume that auditory conditions modulate the physiological responses of a person, \textcolor{red}{we can formulate two predictive tasks, which can be further leveraged to predict the attention score}.

\subsection{\textbf{T2}: Noise level prediction}
\label{ss:T2}
Assuming that the background noise level influences the physiological responses of a person, a model $f_N(\cdot)$ can be obtained to provide an estimation $N^{\prime}$ of the noise level experienced by a subject, based on the features extracted from the physiological responses during a listening segment, %of physiological responses, $N = f(F_r)$, where $R \rightarrow Sg_L \rightarrow F_r$, such that to minimize the expected risk on test data;
\begin{equation}
N^{\prime} = f_N(F_r)
\end{equation}
Since the defined noise level is an ordinal quantity, this predictive task can be formulated as a regression problem with a loss functions similar to the ones suggested for the predictive task \textbf{T1}. Alternatively, this predictive task can be formulated as a classification problems, where each of the six levels of noise constitute different classes.

\subsection{\textbf{T3}: Semanticity of stimulus prediction}
\label{ss:T3}
Assuming that the semanticity of a stimulus $S$ modulates the physiological responses of a subject, a model $f_S(\cdot)$ can be obtained to estimate the semanticity experienced by subject from the set of features $F_R$:
\begin{equation}
S^{\prime} = f_S(F_r)
\end{equation}
Since semanticity is a binary valued auditory condition, this task can be formulated as a classification problem. Choices of loss functions include the cross-entropy or Hinge loss.

\subsection{\textbf{T4}: LWR classification} %Listener task prediction :
\label{ss:T4}
Tasks \textbf{T1}, \textbf{T2} and \textbf{T3} are based on listening segments only and therefore assume that the subject not engaged in any activities other than listening. However, there are scenarios where we might be interested in evaluating the activity that a subject might be engaged in.
%assessing whether for a real-world scenario, a model should evaluate first, if subject is even listening, before going further to estimate the experienced noise level, semanticity and attention score. 
The objective of the \textit{LWR}-classification task is to predict which activity $\mathcal{T}$ is being performed by the subject, considering listening, writing or resting as the candidate activities. Similar to other predictive tasks, this can be modeled as to predict $\mathcal{T}^{\prime}$ for unseen data, as follows
%the task $T$, if subject it listening, writing, or resting from physiological responses, as follow;
\begin{equation}
\mathcal{T}^{\prime} = f_T(F_r)  
\end{equation}
where $f_T(F_r)$ is a model that uses a set of features $F_r$ extracted from a segment of physiological responses. %$R \rightarrow F_r$.  The choice of loss function can be similar to semanticity prediction task.

%The predictive tasks \textbf{T2, T3} and \textbf{T4} can be merged to support the prime objective of experiment 

%The objective of different predictive tasks to support the prime objective 

% $$R \rightarrow Fr$$
% $$\mathcal{T}^{\prime} = f_T(Fr)$$
% $$N^{\prime} = f_N(Fr,\mathcal{T}^{\prime} )$$
% $$S^{\prime} = f_S(Fr, \mathcal{T}^{\prime})$$
% $$A^{\prime} = f_A(Fr,S^{\prime},N^{\prime},\mathcal{T}^{\prime}) = f_A \big( Fr,f_T(Fr),f_N(Fr),f_S(Fr) \big)$$

\section{Feature extraction framework} %for predictive tasks}
\label{S:fExtraction}

\begin{figure}[ht]
     \centering
     \begin{subfigure}[b]{0.49\textwidth}
         \centering
         \includegraphics[width=\textwidth]{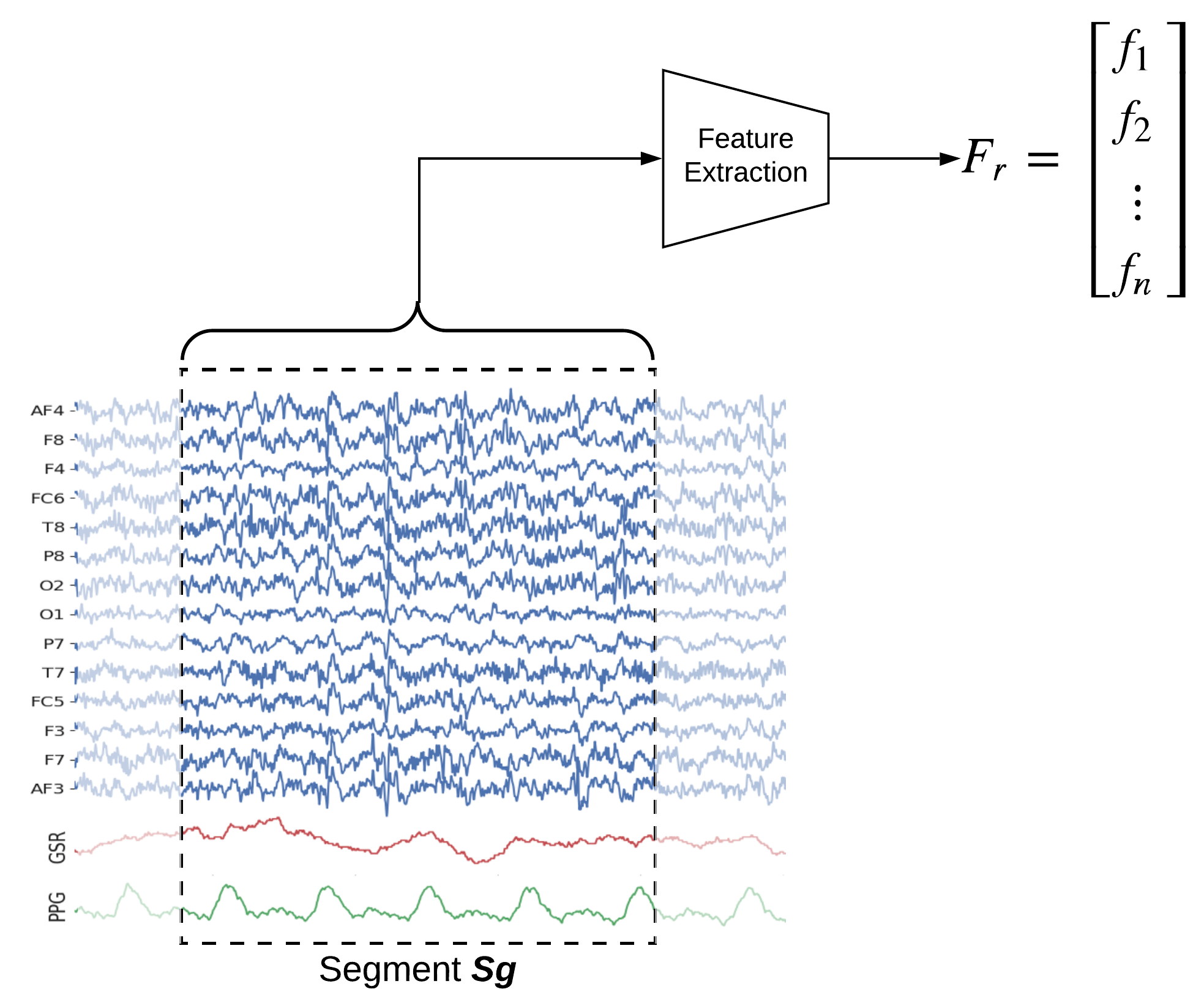}
         \caption{Segment-wise feature extraction}
         \label{fig:fe_segment}
     \end{subfigure}
     \hfill
     \begin{subfigure}[b]{0.49\textwidth}
         \centering
         \includegraphics[width=\textwidth]{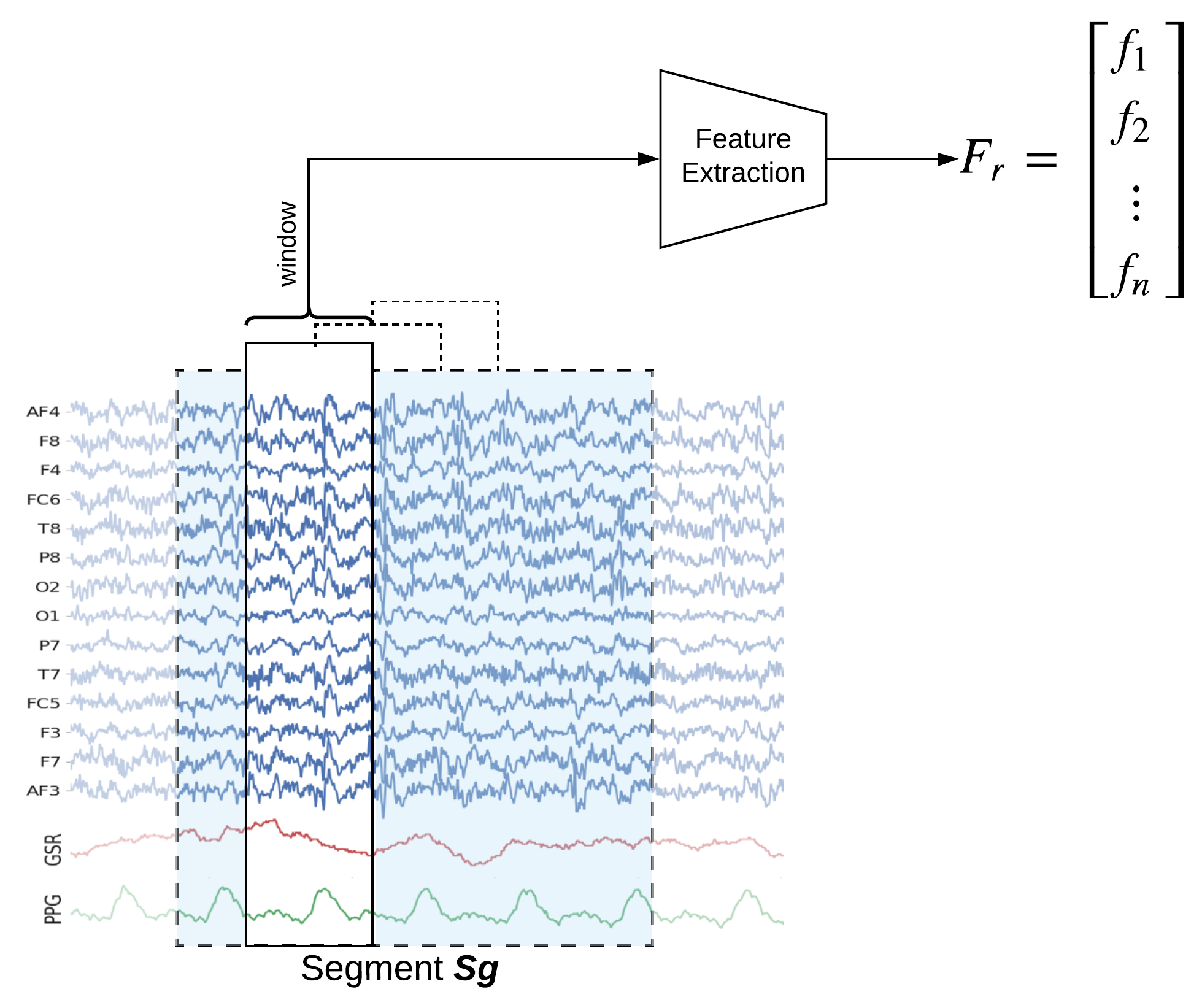}
         \caption{Window-wise feature extraction}
         \label{fig:fe_window}
     \end{subfigure}
        \caption{Feature extraction framework}
        \label{fig:feature_ext}
\end{figure}

In the context of the experiment described in Section \ref{ss:Procedure}, a segment corresponds to a continuous time interval during which a complete task (listening, reading or resting) is executed. By processing the physiological signals recorded during an entire segment, a set of features $F_r$ can be extracted and used as an input in any of the predictive tasks \textbf{T1} through \textbf{T4}. Other time intervals that do not map to complete tasks could also be used to extract features. We will call these time intervals windows, to distinguish them from our definition of segment. A graphical illustration of segment-wise and window-wise feature extraction is shown in Figures  \ref{fig:fe_segment} and \ref{fig:fe_window}, respectively.
%\textcolor{red}{For predictive tasks \textbf{T1} and \textbf{T3}, the target labels; attention score and semanticity are the construct of the entirety of stimulus. In other words, the attention score is computed on the entire transcription of a stimulus and entire stimulus is labeled as semantic or non-semantic. Hence, a valid approach of estimating targets values for \textbf{T1} and \textbf{T3}, should be based on the entire segment. In contrast, for predictive tasks \textbf{T2} and \textbf{T4}, the target labels; noise level and task label, are true for the entire segment as well as for every short duration of the segment. %The level of background noise and the task label are same for the entire segment.
Different models could be built to estimate the target predictions based on features extracted from short windows or from an entire segment. An interesting question is to determine the minimum window length that will produce a set of features $F_r$ that contain enough information to produce an accurate prediction.

%necessary  worth exploring the minimum duration that a model can use to estimate the target value.  %depends on the richness of features and complexity of the model. 

%\textcolor{red}{Since the length of a segment for a task e.g. listening, writing and resting, varies, each predictive task can be handled differently. Based on the task, we define two approaches to extract features. First approach is to extract a fixed set of features $F_r$ from entire segment $Sg$ (Segment-wise), such as spectral features, statistical features etc. Second approach is to extract features from a small, overlapping windows of segment, (window-wise). A pictorial representation of segment-wise and window-wise feature extraction are shown in Figure \ref{fig:fe_segment} and \ref{fig:fe_window}, respectively.}

%\textcolor{red}{The segment-wise feature extraction approach can be used for all the four predictive task. However, 

By allowing moving windows, window-wise feature extraction could be used to train temporal models, such as Dynamic Bayesian Network or RNN, for predictive tasks \textbf{T1} and \textbf{T3}. Features extracted from each window could be modelled as non-Independent and Identically Distributed (IID) samples. In the case of predictive tasks \textbf{T2} and \textbf{T4}, window-wise feature extraction could be used to train temporal or non-temporal models and the set of features extracted from each window could be treated as IID samples instead.

\section{Results of predictive tasks}
\label{S:Results}
%In this section, we discuss the results computed for each predictive task, explained in Section \ref{S:pTasks}, using data of subject \textit{S1}. 
In this section a basic SVM model is created for each of the predictive tasks defined in Section \ref{S:pTasks}, by using data from a single subject, namely \textit{S1}. First the EEG signals from subject S1 are pre-processed by using a high-pass filter with cutoff frequency of $>0.5Hz$ and artifact are removed by using an automatic and tunable wavelet approach \cite{bajaj2020automatic} with default parameter setting, namely $db3, \beta=0.1, N=128$ and $\lambda_s(\cdot)$. After the pre-processing stage,  six spectral features are extracted from each segment from the 14 EEG channels. %Six spectral features for each channel are computed from each segment (segment-wise feature extraction). 
The spectral features correspond to the total power in the commonly used EEG frequency bands,
namely delta (0.5–4 Hz), theta (4–8 Hz), alpha (8–14 Hz), beta (14–30 Hz), low gamma (30–47 Hz) and high gamma (47–64 Hz).%namely; Delta (0.5-4Hz), Theta (4-8Hz), Alpha (8-16Hz), Beta (16-30Hz), Lower Gamma (30-47Hz), and Higher Gamma (47-63.5Hz). 
The spectral power in each frequency band is computed based on Welch's method. In summary, a total of 84 spectral features ($6\times14=84$) are extracted from each EEG segment. 

During the modeling stage, SVM with RBF kernel are used. Task \textbf{T1} and \textbf{T2} are modeled with a Support Vector Regressor. In task \textbf{T2}, noise level $\infty$ dB is mapped to 10 dB. Task \textbf{T3} and \textbf{T4} are modeled with Support Vector Classifier. Mean absolute error (MEA) and accuracy are used as performance metrics for regression (\textbf{T1} \& \textbf{T2}) and classification (\textbf{T3} \& \textbf{T4}), respectively.

\begin{figure}[ht]
    \centering
    \includegraphics[width=0.88\textwidth]{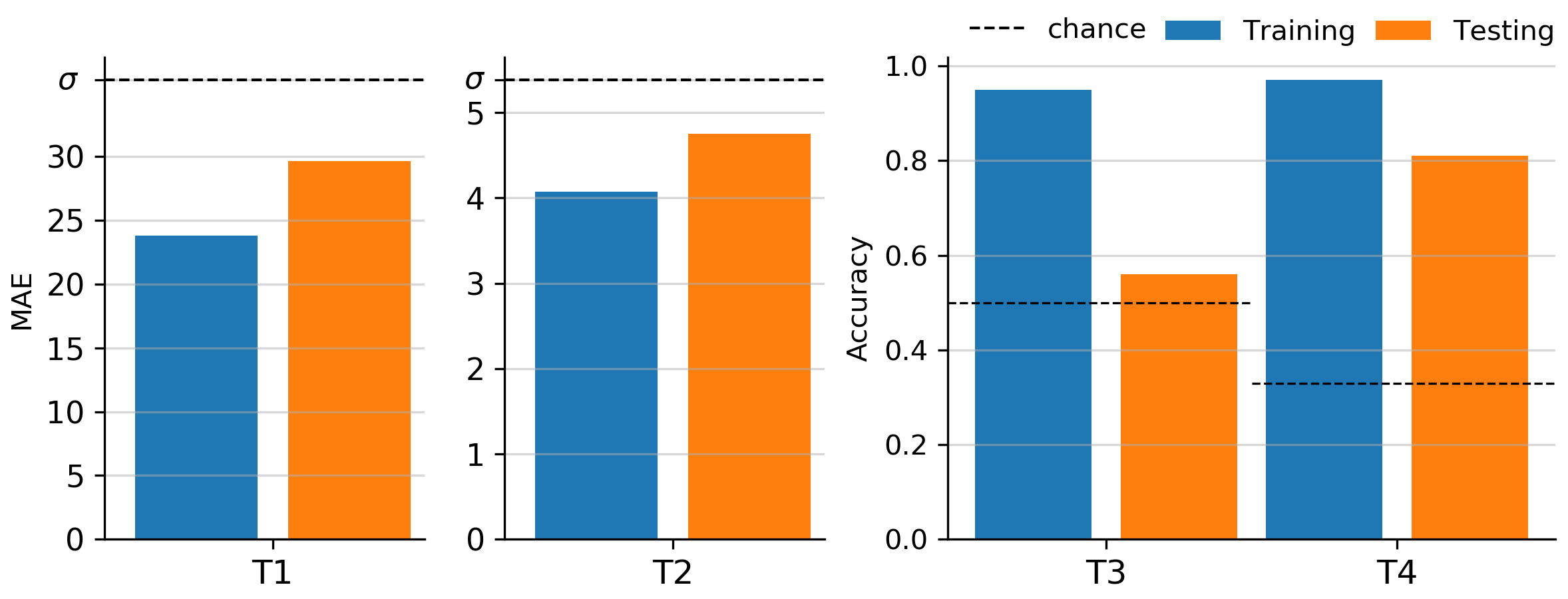}
    \caption{Results of basic modeling using SVM for each predictive task with subject \textit{S1}}
    \label{fig:ptask_results}
\end{figure}

Since the sample size for each task (i.e. $143$ for \textbf{T1, T2, T3} and $143\times3=429$ for \textbf{T4}) is small, K-Fold cross validation with $K=10$ folds is used and to avoid overfitting a regularization parameter $C$ is included. The average training and testing performance for each task together with the final value of the parameter $C$ are reported in Table \ref{table:results} and shown in Figure \ref{fig:ptask_results}. %In order to avoid overfitting, SVM for each task is tuned with regularization parameter $C$ and final value of $C$ is reported. 
It can be observed from Figure \ref{fig:ptask_results} that performance for each predictive task is better than chance level measure. Since the class ratio is balanced for classification tasks, the chance level is 1/number of classes (e.g. 1/2 for \textbf{T3} and 1/3 for \textbf{T4}). For regression, the standard deviation $\sigma$ is shown in Figure \ref{fig:ptask_results}. These results are reproducible and the corresponding python scripts are provided as part of the \textit{phyaat} library. 
\begin{table}[h!]
\caption{Results of basic modeling using SVM for each predictive task with subject \textit{S1}}
\label{table:results}
\centering
\begin{tabular}{l| l|c c |l}
\toprule
\textbf{Predictive Task} &\textbf{SVM} &  Training & Testing & Metric\\
\toprule
T1: Attention score & $C=10$ & 23.80 & 29.65 & \multirow{2}{1.3cm}{MAE}\\[0.1cm]
\cline{1-4}
\noalign{\smallskip}
T2: Noise level & $C=1$& 4.07 & 4.75 & \\
\midrule
T3: Semanticity & $C=2$ & 0.95 & 0.56 & \multirow{2}{1.3cm}{Accuracy}\\[0.1cm]
\cline{1-4}
\noalign{\smallskip}
T4: LWR &$C=2$& 0.97 & 0.81 & \\
\bottomrule
\end{tabular}
\end{table}

\section{Conclusions and Future work}
\label{S:Conclusion}
In this paper, we have presented a dataset of physiological signals for predictive analysis of auditory attention to natural speech. Our dataset is novel in its kind and contains three different physiological signals, namely EEG, GSR, and PPG. %, recorded at sampling rate of 128 Hz. 
Signals were recorded during an experiment conducted with 25 subjects, which involved three tasks (listening, writing and resting) executed in 144 trials . In the listening stage, subjects were presented with an audio stimulus, which is to be transcribed in following writing stage, followed by a resting before the subsequent trial. In each trial, audio stimuli were presented in different auditory conditions, which included background noise level, semanticity, and length of the stimulus. The attention score from transcription of each trial was computed. Based on the collected dataset and experiment design, four predictive tasks have been formulated and two approaches of feature extraction have been proposed. For demonstration purposes, SVM models using EEG spectral features have been created for each predictive tasks using EEG data from a single subject. The correlation analysis of the attention score and the results of the predictive tasks demonstrate the potential of the collected dataset.

In order to advance the research in the field of auditory attention, the collected dataset has been made openly available to the wider research community, together with supporting tools (See Resources section below) %\ref{S:Resources}). 
To improve the performance of the predictive tasks, more sophisticated models such as DBN, RNN and LSTM can be used with spectral, wavelet-based and other features. Our dataset provides an excellent opportunity to investigate the auditory attention mechanisms of the brain. 

\section*{Resources}% - python library - \textit{phyaat}}
\label{S:Resources}
The dataset and supporting source code is distributed as a python library \textit{phyaat} (\href{https://pypi.org/project/phyaat/}{\textcolor{blue}{https://pypi.org/project/phyaat/}}) on the Python Package Index (PyPI) platform. A user guide and scripts to reproduce the results presented in this paper, can be downloaded from the project homepage (\href{https://phyaat.github.io/}{\textcolor{blue}{https://phyaat.github.io/}}). Supporting functions in the library include pre-processing, feature extraction and modeling functions.

\bibliographystyle{unsrt}
\bibliography{references}


%\begin{thebibliography}{1}

%\end{thebibliography}

% \section*{Appendix - Python code}

% % \begin{lstlisting}[language=Python, caption=Python example]
% % pip install phyaat
% % \end{lstlisting}

% \begin{lstlisting}[language=Python, caption=Python example]
% #Install phyaat library
% pip install phyaat

% # import required libraries
% import numpy as np
% import matplotlib.pyplot as plt

% # import phyaat library
% import phyaat as ph

% #download dataset of a subject (subject=1)
% path = ph.loaddata(sub=1,path='../')

% #download dataset of all the subjects
% path = ph.loaddata(sub=-1,path='../')

% # Read file location of all the subjects available
% SubID = ph.Readfiles(path)
% print(SubID.keys())

% #the dataset of subject 1, as segments
% Sub1 = ph.readData(SubID[sub])

% #Apply 1 Hz highpass filter
% Sub1.filter_EEG(type=highpass,cfreq=[1])

% #Apply Artifact Removal Algorithm
% Sub1.Correct(algo='ICA')

% #Read LWR segments
% L,W,R,Labels,col = Sub1.getLWR(onlyEEG=True)

% #get X,y for Task 1, with spectral features, segmentwise
% X = ph.FeatureExtraction(Seg=L,Rhythmic=True,winsize=-1)
% y = Lables['Attention']

% \end{lstlisting}

\end{document}